# What is the Thouless Energy for Ballistic Systems?

Alexander Altland

*Institut für theoretische Physik, Universität zu Köln, Zülpicher Str.77, 50937 Köln, Germany*

Yuval Gefen

*The Weizmann Institute of Science, Department of Condensed Matter Physics, 76100 Rehovot, Israel*

Gilles Montambaux

*Laboratoire de Physique des solides, Université Paris-Sud, 91405 Orsay, France*

The Thouless energy, $E_c$ characterizes numerous quantities associated with sensitivity to boundary conditions in diffusive mesoscopic conductors. What happens to these quantities if the disorder strength is decreased and a transition to the ballistic regime takes place? In the present analysis we refute the intuitively plausible assumption that $E_c$ loses its meaning as an inverse diffusion time through the system at hand, and generally disorder independent scales take over. Instead we find that a variety of (thermodynamic) observables are still characterized by the Thouless energy.

Much of the physics of diffusive disordered conductors is governed by an energy scale $E_c$, known as the Thouless energy [1]. By definition, $E_c = \hbar D/L^2$ ($D$ is the diffusion constant and $L$ the system size) coincides with the inverse of the diffusion time through the system, the latter being a purely classical quantity. Nevertheless, there are numerous energy scales, related to *quantum–mechanical* aspects of disordered conductors, which equal (up to numerical factors) $E_c$ as well:

$E_1$: The disorder averaged curvature (at $\Phi = 0$) of the total energy of an AB-system containing a constant (i.e. flux independent) number of particles. This quantity is proportional to the flux derivative of the (canonical) persistent current $\langle I(\Phi) \rangle_{d,L}$ near zero flux. Here $\langle \ldots \rangle_{d,L}$ denotes impurity and sample size averaging.

$E_2$: The energy scale governing the suppression of the harmonics content $I_m := 2\phi_0^{-1} \int_0^{\phi_0} d\Phi \sin(2\pi m\Phi/\phi_0) I(\Phi)$ of the total persistent current $\langle I(\Phi) \rangle_{d,L}$ ($\phi_0 = h/e$). According to Ref. [2],

$$\langle I_m \rangle_{d,L} = -4\Delta/\phi_0 \exp\left(-2m\sqrt{\gamma/E_c}\right), \qquad (1)$$

where $\gamma$ is an inelastic broadening [3], and $\Delta$ the average level spacing.

$E_3$: The typical current carried by a single electron level (i.e. the derivative of the latter with respect to $\Phi$) averaged over $\Phi$, $\left\langle \overline{i^2(\Phi)} \right\rangle_{d,L} \sim \phi_0^{-2} \Delta E_c$. Here $\overline{(\ldots)}$ refers to flux averaging.

There are further quantities that are not manifestly boundary condition sensitive but nevertheless equal $E_c$ in diffusive systems:

$E_4$: the quantity $g(h/e^2)\Delta$ where $g$ is the average two lead conductance of diffusive conductors [1].

$E_5$: The energy scale limiting the applicability of random matrix theory (RMT) to the description of spectral correlations.

The above list is not exhaustive. Yet, all other occurrences of $E_c$ known to us are closely related to at least one of the examples $E_1, \ldots, E_5$. The realization of the important role played by the Thouless energy can be regarded as one of the major steps towards a comprehensive theory of disordered systems.

The above statements apply to *diffusive* metallic systems, i.e. systems larger than the elastic mean free path $l$. Upon reducing the amount of disorder (or decreasing $L$) a transition to the ballistic regime takes place. This, however, does not imply that disorder becomes immaterial. A simple heuristic argument supporting the opposite may be given as follows: Most of the above mentioned quantities are closely related to the system's quantum-mechanical energy spectra. In a generic $d$-dimensional ballistic system $l > L > l(p_F L)^{1-d}$ ($p_F$ is the Fermi momentum) [4] the typical value of the disorder scattering matrix element exceeds the mean level spacing by a large factor $(p_F L)^{d-1} L/l$ which means that the distribution of the energy eigenvalues differs drastically from the clean case. This implies the existence of a distinct regime (rather than a crossover point), intermediate between the diffusive and the clean limits [5]. The characteristic energy scales in this regime satisfy $\Delta < \tau^{-1} < t_f^{-1} < E_c \ll E_F$, where $\tau$ is the elastic scattering time, $t_f = v_F L^{-1}$ the ballistic time of flight, $v_F$ the Fermi velocity, $E_F$ the Fermi energy, $E_c = v_F l/(d \cdot L^2)$ is defined formally as before and $\hbar$ has been set to unity. The above hierarchy is to be compared with $\Delta < E_c < \tau^{-1} \ll E_F$ characteristic for the diffusive regime.

The existence of strong spectral correlations suggests that at least some of the scales $E_1, \ldots, E_5$ continue to be disorder sensitive even in the ballistic regime. On the other hand, $E_c$ loses its meaning as an inverse transport time through the system, in which respect it is replaced by the - disorder independent - inverse time of flight $t_f$. It is thus imperative to ask whether it is possible to find an energy scale in the ballistic regime playing role analogous and as crucial as $E_c$ in the diffusive case and whether this scale is disorder dependent. The results of our analysis, reported here, show that for boundary condition (or magnetic flux) sensitive quantities *the Thouless energy, $E_c$, is still the relevant scale.*

To be specific, we consider an ensemble of two-dimensional AB cylinders of height and circumference $L$ which differ in their impurity configuration and size. The relevance of sample size averaging will become evident below. Apart from the average conductance, $E_4$, all the above listed quantities can be calculated by means of the correlation function

$$K^{\text{ens}}(\epsilon,\phi_1,\phi_2) := \frac{1}{\nu_0^2}\langle \nu(E_F+\epsilon,\phi_1)\nu(E_F,\phi_2)\rangle_{L,d} - 1$$

characterizing fluctuations in the energy spectra of the individual systems. Here $\phi := 2\pi\Phi/\phi_0$, $\nu(\epsilon,\phi)$ is the flux dependent densitiy of states and $\nu_0 := \langle\nu(\epsilon,\phi)\rangle_{L,d}$ is practically constant. In order to evaluate this expression in the particular case of a ballistic ensemble, it is essential to note two features of the combined disorder/sample size average: (i) Size and disorder average are essentially statistically independent, $\langle\ldots\rangle_{L,d} \simeq \langle\langle\ldots\rangle_d\rangle_L$ [4] and (ii) size averaging is equivalent to averaging over the Fermi-energy (see below). For our purposes it suffices to consider an average over a range $E_0 > \Delta_1 \simeq (p_F L)^{d-1}\Delta$ corresponding to size fluctuations on scales $\sim 1/p_F$. From (i) and (ii) it follows immediately that [11]

$$K^{\text{ens}}(\epsilon,\phi_1,\phi_2) = \langle K(\epsilon,\phi_1,\phi_2)\rangle_{E_0} + K'(\epsilon,\phi_1,\phi_2), \quad (2)$$

where

$$K'(\epsilon,\phi_1,\phi_2) := \frac{1}{\nu_0^2}\langle\langle\nu(E_F+\epsilon,\phi_1)\rangle_d\langle\nu(E_F,\phi_2)\rangle_d\rangle_{E_0} - 1,$$

$$K(\epsilon,\phi_1,\phi_2) := \frac{1}{\nu_0^2}\Big(\langle\nu(E_F+\epsilon,\phi_1)\nu(E_F,\phi_2)\rangle_d -$$

$$-\langle\nu(E_F+\epsilon,\phi_1)\rangle_d\langle\nu(E_F,\phi_2)\rangle_d\Big), \quad (3)$$

In the diffusive regime, $\langle\nu(\epsilon,\phi)\rangle_d \simeq \nu_0 \Rightarrow K' \simeq 0$ implying that the structure of the spectrum is solely described disorder induced correlation function $K$. On the other hand, it is clear from the defining equations that $K \to 0$ in the clean limit, in which case the function $K'$, corresponding to correlations in the clean spectrum, gains importance. This motivates the above definition of the ensemble average: The structure of the clean spectrum depends sensitively on various parameters, such as the system size or the position of the energy window under consideration. Any realistic modelling of an ensemble average should thus involve a simultaneous average over both, disorder and these latter characteristics. In an ensemble of diffusive systems the fluctuations induced by disorder are strong enough to mask any underlying clean structure and averaging over other parameters is inessential. In the course of a crossover to the clean limit the structure of the clean spectrum becomes visible and the function $K'$ begins to play a dominant role. The ballistic case represents a transition regime in the sense that the functions $K$ and $K'$ are simultaneously relevant. We next discuss both contributions separately.

Within the framework of diagrammatic perturbation theory, the *disorder induced correlation function $K$* [4] can be written as the sum of a ladder type diagram and its time reversed analogue [3], here referred to as the diffuson and the Cooperon contribution, respectively. In the time reversal invariant (field free) case, both diagrams are equal and one arrives at $K(\epsilon,0,0) = \Delta^2/(2\pi)^2 \text{Re } \partial_\epsilon^2 2\sum_q \hat{\ln}(1-\lambda(\vec{q},\epsilon)) + \ldots$, where $\sum_q$ denotes a summation over a two-dimensional vector $\vec{q} = (2\pi n_x/L_x, \pi n_y/L_y)$, $n_x \in \mathcal{Z}, n_y \in \mathcal{N}$ of quantized modes, $\lambda(\vec{q},\epsilon) = \left[(1+\gamma\tau+i\epsilon\tau)^2 + (|\vec{q}|l)^2\right]^{-1/2}$, $\gamma \simeq \Delta$ is an inelastic broadening, $\hat{\ln}(1+z) := \ln(1+z) - z$ and the dots indicate small and field insensitive corrections which may be discarded in the present discussion.

Upon the inclusion of magnetic fluxes, the degeneracy between the diffuson and the Cooperon contribution is lifted and we obtain $K(\epsilon,\phi_1,\phi_2) = \Delta^2/(2\pi)^2 \text{Re } \partial_\epsilon^2 F(\epsilon,\phi_1,\phi_2)$, where $F(\epsilon,\phi_1,\phi_2) = \pi^{-2}\sum_q \left(\hat{\ln}\left[1-\lambda\left(\vec{q}+L^{-1}\phi_+\vec{e_x},\epsilon\right)\right] + (\phi_+ \leftrightarrow \phi_-)\right)$, where $\phi_\pm := \phi_1 \pm \phi_2$, $\vec{e_x}$ is the unit vector in $x$-direction and $\phi_+$ ($\phi_-$) corresponds to the Cooperon (diffuson) contribution. For practical convenience we next introduce the function $\langle\delta N^2(\phi)\rangle^{\text{ens}}$ describing the flux sensitive contribution to the fluctuation of the total number of occupied levels. This quantity relates to the above introduced function $K^{\text{ens}}$ via $\langle\delta N^2(\phi)\rangle^{\text{ens}} = \Delta^{-2}\int_0^{E_F} d\epsilon_1 d\epsilon_2 \left[K^{\text{ens}}(\epsilon_1-\epsilon_2,\phi,\phi) - K^{\text{ens}}(\epsilon_1-\epsilon_2,0,0)\right]$. In analogy to Eq. (2) we further define $\langle\delta N^2(\phi)\rangle^{\text{ens}} =: \langle\delta N^2(\phi)\rangle + \langle\delta N^2(\phi)\rangle'$, where the primed (unprimed) term corresponds to the $K'$ ($\langle K\rangle_{E_0}$) contribution to $K^{\text{ens}}$. Due to the fact that $K$ has been expressed in terms of a twofold derivative, the energy integration necessary to obtain $\langle\delta N^2(\phi)\rangle$ can trivially be performed and we arrive at $\langle\delta N^2(\phi)\rangle = F(E_F,\phi,\phi) - F(0,\phi,\phi) - F(E_F,0,0) + F(0,0,0)$. An approximate harmonic expansion of $\langle\delta N^2(\phi)\rangle$ reads

$$\langle\delta N^2(\phi)\rangle \simeq \frac{4}{\pi^2}\sum_{m=1}^\infty \left(\frac{1}{m}e^{-2m\sqrt{\gamma/E_c}} - Ae^{-2mB/\sqrt{E_c\tau}}\right) \cdot$$
$$\cdot [\cos(2m\phi) - 1], \qquad (4)$$

where $A, B$ are constants of $\mathcal{O}(1)$ (Within our numerical accuracy we obtain $A \simeq 0.92$, $B \simeq 1.21$).

We next discuss the *'clean' correlation function $K'$*. Employing the Poisson sum rule to account for the discreteness of the energy levels, one may write the disorder averaged density of states as

$$\langle \nu(E,\phi)\rangle_d = \frac{1}{2\pi^2}\sum_{k\in\mathcal{Z}\times\mathcal{N}}\int d^2p\, e^{i\vec{p}\cdot\vec{L}_k}\,\mathrm{Im}\,G^-(p,E,\phi) =$$

$$= \frac{1}{2\pi^2}\sum_{k\in\mathcal{Z}\times\mathcal{N}}\int d^2p\, e^{i\vec{p}\cdot\vec{L}_k} e^{iL\phi}\mathrm{Im}\,G^-(p,E,0), \quad (5)$$

where $\sum_{k\in\mathcal{Z}\times\mathcal{N}}$ is an abbreviation for the summation over all pairs $k = (k_x, k_y)$, $k_x \in \mathcal{Z}, k_y \in \mathcal{N}$, $\vec{L}_k^T = L(k_x, 2k_y)$ and $G^{-1}(p,E,\phi) = \left(E - \frac{i}{2\tau} - \frac{1}{2m}(\vec{p} - \frac{1}{L}\phi\vec{e_x})^2\right)^{-1}$. Substituting this expression in the definition, Eq. (3), we obtain

$$K'(\epsilon,\phi_1,\phi_2) = \frac{1}{2\pi p_F}\sum_{k_1,k_2\in M}\frac{e^{-\frac{L_{k_1}+L_{k_2}}{2l}}}{(L_{k_1}L_{k_2})^{1/2}} \times$$

$$\times \left\langle e^{iL_{k_1}p(E_F+\epsilon) - iL_{k_2}p(E_F)} + c.c.\right\rangle_{E_0} e^{i(k_{1,x}\phi_1 + k_{2,x}\phi_2)}, \quad (6)$$

where $L_{k_i} = |\vec{L}_{k_i}|$, $p(E) = \sqrt{2mE}$ and $M = \mathcal{Z}\times\mathcal{N} - (0,0)$. From the structure of the expression appearing under the angular brackets, it is clear that size and energy averaging are equivalent: Both procedures eliminate contributions oscillating like $\exp(i\cdot\mathrm{const.}\cdot p_F L)$. Hence we obtain

$$K'(\epsilon,\phi_1,\phi_2) = \frac{1}{2\pi p_F}\sum_{k\in M}\frac{1}{L_k}\cos\left(\epsilon\frac{L_k}{v_F}\right) \times$$

$$\times e^{-L_k/l}\left[\cos(k_x\phi_+) + \cos(k_x\phi_-)\right].$$

This expression immediately leads to

$$\langle\delta N^2(\phi)\rangle' = \frac{2v_F^2}{\pi p_F\Delta^2}\sum_{k\in M}\frac{e^{-L_k/l}}{L_k^3}\left[\cos(2k_x\phi) - 1\right].$$

Approximating the summation over the component $k_y$ by an integration, we finally find

$$\langle\delta N^2(\phi)\rangle' = \frac{2}{\pi}p_F L\sum_{m=1}^{\infty}\frac{e^{-2mL/l}}{m^2}\left[\cos(2m\phi) - 1\right]. \quad (7)$$

Technically, Eqs. (4) and (7) represent the central result of this Letter. We next apply it to the discussion of the energy scales $E_1,\ldots,E_5$.

The first two scales $E_1$ and $E_2$ are directly related to the ensemble averaged persistent current (of non–interacting electrons!) [2]

$$\langle I(\phi)\rangle_{d,L} = -\frac{\Delta}{2}\frac{2\pi}{\phi_0}\partial_\phi\langle\delta N^2(\phi)\rangle \quad (8)$$

$E_1$: The curvature of the total energy at zero flux, i.e. the slope of the total persistent current is obtained by differentiating $\langle\delta N^2(\phi)\rangle^{\mathrm{ens}}$ twice with respect to flux (cf. Eq. (8)) [10]. As follows from Eq. (4), $-\frac{\Delta}{2}\left(\frac{2\pi}{\phi_0}\right)^2\partial_\phi^2|_{\phi=0}\langle\delta N^2(\phi)\rangle \sim \frac{\Delta}{\gamma}E_c$, whereas the analogous contribution of $\langle\delta N^2(\phi)\rangle'$ is given by (cf. Eq. (7))

$\sim \Delta p_F L \sum_{m=1}^{l/L}\frac{1}{m^2}m^2 \sim E_c$. We thus conclude that $E_1$ is equal to $E_c$.

$E_2$: The harmonic content of the current, is obtained by inserting the decomposition $\langle\delta N^2(\phi)\rangle^{\mathrm{ens}} = \langle\delta N^2(\phi)\rangle + \langle\delta N^2(\phi)\rangle'$ into Eq. (8). Due to the large factor $p_F L \gg 1$ appearing in Eq. (7) the behaviour of the low harmonics $m < L/l$ is dominated by the $\langle\delta N^2(\phi)\rangle'$–contribution. For $m > L/l$, however, the latter is exponentially suppressed and the harmonics take the same form as in the diffusive regime (cf. Eq. (1)). In particular, the scale governing the overall exponential suppression of the harmonic content is given by $E_c$.

$E_3$: Employing the thermodynamic relation $\frac{\partial I(\phi,\mu)}{\partial \mu} = \frac{\partial N(\Phi,\mu)}{\partial \Phi}$ one may evaluate the typical single level current (in the vicinity of $E_F$) by computing $\left\langle i^2(\phi)\right\rangle_{d,L} = \left(\frac{2\pi}{\phi_0}\right)^2\partial^2_{\phi_1,\phi_2}\int_0^{E_F}d\epsilon_1 d\epsilon_2 K^{\mathrm{ens}}(\epsilon_1 - \epsilon_2,\phi_1,\phi_2)\Big|_{\phi_1=\phi_2}$. This quantity can straightforwardly be obtained by generalizing above derivation to the case of non–coinciding fluxes $\phi_1$ and $\phi_2$. As a result one obtains generalizations of $\langle\delta N^2(\phi)\rangle$ respectively $\langle\delta N^2(\phi)\rangle'$ in which the characteristic factor $\cos(2\phi) - 1$ is replaced by $\cos(\phi_+) + \cos(\phi_-) - 2$. Consequently we find

$$\left\langle\overline{i^2(\phi)}\right\rangle_{d,L} = -\frac{\Delta}{4}\frac{2\pi}{\phi_0}\frac{\partial\langle I(\phi)\rangle_{L,d}}{\partial\phi}\Big|_{\phi=0},$$

leading to $E_3 = E_c$.

The relevance of the Thouless energy for $E_1, E_2$ and $E_3$ is most easily understood from a semiclassical point of view: All these scales are governed by the contribution of high harmonics $m \gg 1$ to the sum Eq. (4). The behaviour of any harmonic $m$ is determined by Feynman–paths winding typically $m$-times around the ring. As soon as $m > l/L$, the motion along these paths is effectively 'diffusive' due to disorder scattering [11] and we obtain a corresponding result for the associated energy scales.

Compared to these flux–dependent thermodynamic observables, flux insensitive or transport observables change much more drastically in the course of a transition to the ballistic regime. In all cases known to us, $E_c$ loses its meaning and other energy scales begin to dominate. As an example we consider the two quantities $E_4$ and $E_5$.

$E_4$: The average two-lead conductance of a ballistic wire is given by the number of conducting channels leading to $E_4 \sim t_f^{-1}$.

$E_5$: It has been found in Ref. [4] that the *disorder induced* spectral correlation functions obey (orthogonal) Wigner dyson statistics as long as $\epsilon < \tau^{-1}$. We thus conclude $E_5 = \tau^{-1}$ (ergodicity on time scales $t > \tau$).

In the above examples of flux dependent observables we have stressed analogies between diffusive and ballistic systems and emphasized the importance of disorder scattering. On the other hand, it is clear that *all* kinds of observables must in some or the other way approach their 'clean' value the more the disorder concentration is reduced. In the present context it is the newly introduced function $K'$ which guarantees the existence of a proper clean limit. To illustrate this, we consider a final example, namely the *amplitude* of the average persistent current.

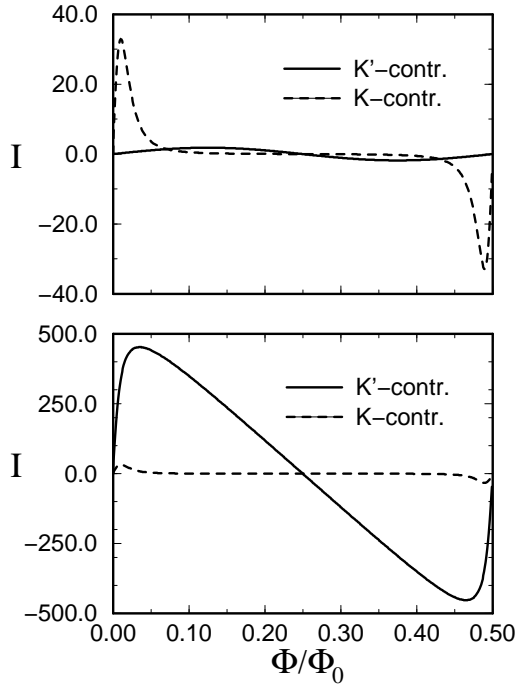

FIG. 1. Average persistent current $\langle I(\phi)\rangle_{d,L}$ vs. flux ($p_F l = 1000$).

The behaviour of the persistent current in the transition region between the diffusive and the ballistic regime has previously been analyzed numerically [7] but defied analytical studies so far. Earlier analytical results were obtained either in the strongly disordered or in the nearly clean limit but disagreed by a large parameter $\mathcal{O}(p_F l)$, when evaluated at the boarderline $L = l$. Upon characterizing the spectrum in terms of the two functions $K$ and $K'$ this discrepancy disappears and one may smoothly interpolate between both limiting cases: As long as $L > l$, the $\langle \delta N^2(\phi)\rangle'$-contribution to the current is exponentially suppressed (cf.fig) and we obtain the known [2,6] result for the persistent current, $\langle I(\phi)\rangle_{d,L}^{\text{diff.}}$ in diffusive rings. Note that $\langle I(\phi)\rangle_{d,L}^{\text{diff.}} \sim E_c \phi$ increases linearly until it reaches its maximum value $I_{\max}^{\text{diff.}} \sim (\Delta E_c)^{1/2}$ at $\phi \sim (\Delta/E_c)^{1/2}$. Upon decreasing the ratio $L/l$, the $\langle \delta N^2(\phi)\rangle'$ term gains importance: In the transition region $L \simeq l$, its contribution is comparable to the $\langle \delta N^2(\phi)\rangle$-term and for $L < l$ it dominates (cf.fig.). Like in the diffusive case, the persistent current carried by ballistic rings, $\langle I(\phi)\rangle_{d,L}^{\text{bal.}}$ sets out linearly $\sim E_c \phi$ at zero flux. The linearly increasing behaviour, however, persists up to larger flux values $\phi \sim \frac{L}{l} \gg (\Delta/E_c)^{1/2}$ leading to a maximum value $I_{\max}^{\text{bal.}} \sim v_F/L$ which exceeds $I_{\max}^{\text{diff.}}$ by the factor $(p_F L)^{1/2}(L/l)^{1/2} \gg 1$. For flux values $\phi > \frac{L}{l}$ the $\langle I(\phi)\rangle_{d,L}^{\text{bal.}}$-curve bears resemblance with the sawtooth like shape obtained for clean rings [9].

Summarizing, we find that ballistic disordered systems behave to some extent 'diffusive' in their static response to external magnetic fluxes. In particular the scale $E_c$, defined formally like in the diffusive regime, remains the cahracteristic energy for various observables. This is notwithstanding the fact that the $E_c$ in the ballistic regime does no longer carry the appealing significance of an inverse transport time through the system. When considering physical quantities of different nature, the overall picture in the ballistic regime turns out to be more complex than in the diffusive case.

*Acknowledgements:* Y.G. acknowledges the hospitality of H.Bouchiat and G.Montambaux in Orsay. This work was supported by the German-Israel Foundation (GIF), the U.S.-Israel Binational Science Foundation (BSF) and the EC Science Program under grant number SSS-CT90-0020.